\documentclass[letterpaper,10pt]{article}
\pdfoutput=1 
\usepackage{opticameet3} %% use version 3 for proper copyright statement

\usepackage[mathscr]{eucal}
\usepackage{subcaption}
\usepackage{adjustbox}
\usepackage{tikz}
\usepackage{pgfplots}
\usetikzlibrary{patterns}
\usepgfplotslibrary{fillbetween}
\usetikzlibrary{positioning,shadows,shapes,backgrounds,calc,spy,arrows}
\usepackage{pgfplotstable}
\pgfplotsset{compat=newest}
\usepackage{tcolorbox}
\usepackage{multicol}
\usepackage{float}
\usepackage{wrapfig}
\usepackage{lipsum}
\usepackage{ragged2e}

\newcommand\authormark[1]{\textsuperscript{#1}}
\newcommand{\thn}{\mathrm{th}}
\newcommand{\rin}{\mathrm{rin}}
\newcommand\notsotiny{\@setfontsize\notsotiny\@vipt\@viipt}

\usepackage{amsmath,amssymb}
\usepackage[colorlinks=true,bookmarks=false,citecolor=blue,urlcolor=blue]{hyperref} %pdflatex

\begin{document}

\title{On Geometric Shaping for 400 Gbps IM-DD Links with Laser Intensity Noise}
%\textcolor{purple}{[v2]}
%\title{On the Effect of Constellation-Shaped PAM for Theoretical 400Gb/s per Lane IM-DD Links}

% \author{Author name(s)}
% \address{Author affiliation and full address}
% \email{e-mail address}
%%Uncomment the following line to override copyright year from the default current year.
\copyrightyear{2025}

\author{Felipe~Villenas\authormark{(1,*)}, Kaiquan~Wu\authormark{(1)}, Yunus~Can~G\"{u}ltekin\authormark{(1)}, Jamal~Riani\authormark{(1,2)}, Alex~Alvarado\authormark{(1)}}

%\address{\authormark{1} Information and Communication Theory Lab, Eindhoven University of Technology, Eindhoven, The Netherlands\\
\address{\authormark{1} Department of Electrical Engineering, Eindhoven University of Technology, Eindhoven, The Netherlands\\
\authormark{2}Marvell Technology, Santa Clara, CA, USA}

\email{\authormark{*}f.i.villenas.cortez@tue.nl} %% email address is required

\begin{abstract}
We propose geometric shaping for IM-DD links dominated by relative intensity noise (RIN). For 400 Gbps links, our geometrically-shaped constellations result in error probability improvements that relaxes the RIN laser design by 3 dB.
%We propose geometric shaping for 400 Gbps IM-DD links dominated by relative intensity noise (RIN). Our geometrically-shaped constellations result in an improvement in SER that can relax the RIN laser design constraints by 3 dB.
%We study geometric shaping (GS) for 400 Gbps intensity modulation and direct detection (IM-DD) links dominated by relative intensity noise (RIN). Our geometrically-shaped constellations result in an improvement in symbol error rates (SERs) that can relax the RIN laser design constraints by 3 dB.
\end{abstract}

\section{Introduction}

There has been a constant need for increasing the data rates in short-reach data center interconnects (DCIs), mainly due to the exploding demand from artificial intelligence applications. Mainstream DCI solutions typically employ optical links with intensity modulation (IM) and direct detection (DD) using pulse amplitude modulation (PAM) formats. This combination enables cost-effective, low-complexity, and low-power transceivers \cite{che2023modulation}.

Most of current 200 Gbps IM-DD systems utilize PAM-4. Higher data rates, such as 400 Gbps or beyond, are mostly restricted by the constraints imposed by the bandwidth of the electronic components of the transceivers, as well as noise tolerance of modulation formats beyond PAM-4. In order to alleviate the electrical bandwidth requirements, PAM-6 and PAM-8 are being considered for 400 Gbps links \cite{hossain2021single}. 

\begin{figure}[H]
    \centering
    \begin{subfigure}[b]{0.48\textwidth}
        \begin{adjustbox}{width=0.92\linewidth,trim={0.2cm 0 0 0}, clip}
            \definecolor{my_green}{rgb}{0.25, 0.6, 0.0}    %en la variables "my_green" guardo un color que he definido a partir de las componentes RGB. Es recomendable definir colores porque las librerías de tikZ no suelen tener mucha variedad de colores.
\definecolor{my_purple}{rgb}{0.5, 0, 0.5}
\definecolor{azure}{rgb}{0, 0.5, 0.5}

\newcommand{\lw}{0.8pt}

\newcommand{\graysolid}{\raisebox{2pt}{\tikz{\draw[color=black!50,solid,line width=\lw,mark=o,mark options={solid}](0,0) -- (2.5mm,0);}}}   %Símbolo de curva solida
\newcommand{\redsolid}{\raisebox{2pt}{\tikz{\draw[color=red,solid,line width=\lw,mark=o,mark options={solid}](0,0) -- (2.5mm,0);}}}   %Símbolo de curva solida
\newcommand{\bluesolid}{\raisebox{2pt}{\tikz{\draw[color=blue,solid,line width=\lw,mark=o,mark options={solid}](0,0) -- (2.5mm,0);}}}   %Símbolo de curva solida
\newcommand{\purplesolid}{\raisebox{2pt}{\tikz{\draw[color=my_purple,solid,line width=\lw,mark=o,mark options={solid}](0,0) -- (2.5mm,0);}}}   %Símbolo de curva 
\newcommand{\greensolid}{\raisebox{2pt}{\tikz{\draw[color=my_green,solid,line width=\lw,mark=o,mark options={solid}](0,0) -- (2.5mm,0);}}}   %Símbolo de curva solida

\newcommand{\blacksdashed}{\raisebox{2pt}{\tikz{\draw[color=black!70,densely dashed,line width=1.0pt,mark=o,mark options={solid}](0,0) -- (2.5mm,0);}}}   %Símbolo de curva discontinua
\newcommand{\cyansolid}{\raisebox{2pt}{\tikz{\draw[color=cyan,solid,line width=1.0pt,mark=o,mark options={solid}](0,0) -- (2.5mm,0);}}}   %Símbolo de curva 

\begin{tikzpicture}
    \begin{axis}[
    width=\linewidth,  %ancho de la figura
    height=1.5in,
    xmin=-7, xmax=7,
    ymin=0, ymax=1.3,
    xlabel={Channel Observations $Y$},
    x label style={yshift=3pt},
    xtick={-5,-3,...,5},
    yticklabels={\empty},
    ylabel={Relative Frequency},
    y label style={yshift=-5pt},
    %ybar,
    ytick pos=left,
    xtick pos=bottom,
    font=\footnotesize,
    minor tick num=1,
    grid style = {dotted,lightgray},
    legend style = {legend pos=north east, font=\footnotesize, legend cell align=left, row sep=-0.5ex},
    domain=-7:7,
    samples=701,
    %ytickten={-8,-6,...,0},
    %xtick={2.7749, 4.7749, 6.7749, 8.7749},
    %xticklabels={$x_0$,$x_1$,$x_2$,$x_3$},
    ]

        \pgfplotstableread{FigTikz/data_txt/histogram_rx_symbols_4a.txt}\datatable
        
        %\addplot[color=black!50, solid, line width=\lw] table[x={oma},y={awgn}] {FigTikz/data_txt/SER_PAM4.txt};
                
        % RIN
        \addplot [draw=none,fill=red,fill opacity=0.5,ybar interval,mark=no] table[x index=0, y index=1] {\datatable};
        \addplot [draw=none,fill=blue,fill opacity=0.5,ybar interval,mark=no] table[x index=2, y index=3] {\datatable};
        \addplot [draw=none,fill=my_green,fill opacity=0.5,ybar interval,mark=no] table[x index=4, y index=5] {\datatable};
        \addplot [draw=none,fill=orange,fill opacity=0.5,ybar interval,mark=no] table[x index=6, y index=7] {\datatable};
        \addplot [draw=none,fill=my_purple,fill opacity=0.5,ybar interval,mark=no] table[x index=8, y index=9] {\datatable};
        \addplot [draw=none,fill=azure,fill opacity=0.5,ybar interval,mark=no] table[x index=10, y index=11] {\datatable};
        %\addplot [color=red!50,fill=red,ybar interval,mark=no] file {FigTikz/data_txt/histogram_data.txt};
        %\addlegendentry{$-140$};
        \foreach \sm in {-5,-3,...,5} {
            \addplot [cyan, solid, line width=1.0pt] {exp(-(x-\sm)^2/(2*(0.0407 + (\sm+16.26)^2*8.3263e-4) ))/sqrt(2*pi*(0.0407 + (\sm+16.26)^2*8.3263e-4))};
        }

        % AWGN
        \foreach \sm in {-5,-3,...,5} {
            \addplot [black!70, densely dashed, line width=1.0pt] {exp(-(x-\sm)^2/(2*0.1463))/sqrt(2*pi*0.1463)};
        }
        
        \node [draw,font=\tiny,fill=white,anchor=north east] at (rel axis cs:1,1) {\shortstack[l]{
        \blacksdashed \hspace{0.01cm} AWGN \hspace{0.025cm} \cyansolid \hspace{0.01cm} AWGN+RIN
        }};

        % Symbols
        \node [font=\scriptsize] at (axis cs:-5,0.15) {$x_0$};
        \node [font=\scriptsize] at (axis cs:-3,0.15) {$x_1$};
        \node [font=\scriptsize] at (axis cs:-1,0.15) {$x_2$};
        \node [font=\scriptsize] at (axis cs:1,0.15) {$x_3$};
        \node [font=\scriptsize] at (axis cs:3,0.15) {$x_4$};
        \node [font=\scriptsize] at (axis cs:5,0.15) {$x_5$};

        % Thresholds {-4.0615   -2.0565   -0.0519    1.9523    3.9559}
        \draw[red, densely dotted, line width=0.8pt] (-4.0615,0) -- (-4.0615,0.75);
        \node [red, font=\scriptsize] at (axis cs:-4.0615,0.85) {$x_0^{\mathrm{th}}$};
        \draw[red, densely dotted, line width=0.8pt] (-2.0565,0) -- (-2.0565,0.75);
        \node [red, font=\scriptsize] at (axis cs:-2.0565,0.85) {$x_1^{\mathrm{th}}$};
        \draw[red, densely dotted, line width=0.8pt] (-0.0519,0) -- (-0.0519,0.75);
        \node [red, font=\scriptsize] at (axis cs:-0.0519,0.85) {$x_2^{\mathrm{th}}$};
        \draw[red, densely dotted, line width=0.8pt] ( 1.9523,0) -- ( 1.9523,0.75);
        \node [red, font=\scriptsize] at (axis cs: 1.9523,0.85) {$x_3^{\mathrm{th}}$};
        \draw[red, densely dotted, line width=0.8pt] (3.9559,0) -- (3.9559,0.75);
        \node [red, font=\scriptsize] at (axis cs:3.9559,0.85) {$x_4^{\mathrm{th}}$};
    \end{axis}

\end{tikzpicture}       % PAM-6
        \end{adjustbox}
    \end{subfigure}
    ~
    \begin{subfigure}[b]{0.48\textwidth}
        \begin{adjustbox}{width=\linewidth,trim={0.5cm 0 0.2cm 0}, clip}
            \definecolor{ashgrey}{rgb}{0.75, 0.75, 0.75}
\definecolor{antiquebrass}{rgb}{0.98, 0.81, 0.69}
\definecolor{brilliantlavender}{rgb}{0.96, 0.73, 1.0}

\tikzstyle{block} = [draw, line width = 1pt, fill=black!20, rectangle, minimum height=30pt, rounded corners=0.1cm, text width=2.5em,align=center]

\tikzstyle{block_wide} = [draw, line width = 1pt, fill=black!20, rectangle, minimum height=30pt, rounded corners=0.1cm, text width=3.5em,align=center]

\tikzstyle{block_wide2} = [draw, line width = 1pt, fill=black!20, rectangle, minimum height=20pt, rounded corners=0.1cm, text width=3em,align=center]

\tikzstyle{block2} = [draw, line width = 1pt, fill=black!20, rectangle, minimum height=30pt, minimum width=30pt, rounded corners=0.1cm, text width=5em,align=center]

\tikzstyle{Cir} = [draw, circle,  minimum size=2.15em]

\begin{tikzpicture}[auto, node distance=1 cm,>=to,line width=0.5pt]

    %%%% IM/DD
    %% placing the blocks
    % TX 
    \node [coordinate] (input) {};  
    \node [block, right = 1.5em of input, fill = antiquebrass] (DAC) {DAC};  
    \node [block,right = 2em of DAC, fill = green!20] (EO) {MZM};
    \node[below=0em of EO](){$\beta$};   
    %\node [block_wide2,above = 1.5em of EO, fill = brilliantlavender!40] (Laser) {Laser};
    \node [block_wide2,above = 1.5em of EO, fill = red!20] (Laser) {Laser};
    \node[above=0em of Laser](){RIN};

    % Fiber
    \draw [color=blue,solid,line width=0.5pt,opacity=1] ($(EO.east)+(0.9,0.3)$) circle (3mm);
    \draw [color=blue,solid,line width=0.5pt,opacity=1] ($(EO.east)+(1.0,0.3)$) circle (3mm);
    \draw [color=blue,solid,line width=0.5pt,opacity=1] ($(EO.east)+(1.1,0.3)$) circle (3mm);
    \node[right=0.5cm of EO, yshift=-1.7em](){$e^{-\alpha L}$};
    
    % % RX
    \node [block_wide, right = 2cm of EO, fill = green!20] (OE) {PD-TIA};
    \node[below=0em of OE](){$\mathfrak{R}, G, B$};
    \node[above=1.8em of OE](noise){AWGN};
    \node [block, right = 2em of  OE, fill = antiquebrass ] (ADC) {ADC};
    \node [right = 1.5em of ADC] (output) {};

    % % We draw edges between nodes
    \draw [draw,-latex] (input) -- node[left,text width=1.2em, align = left]{$X$}(DAC);
    % \draw [draw,-latex] (input) -- node[left]{$X$} (OOK);
    \draw [draw,-latex] (DAC) -- (EO);
    \draw [draw,-latex, color=blue] (Laser) -- (EO);
    \draw [draw,-latex, color=blue] (EO) -- node[midway,below]{\textcolor{black}{Fiber}}(OE);
    \draw [draw,-latex] (noise) -- (OE);
    \draw [draw,-latex] (OE) -- (ADC);
    % \draw [draw,-latex] (INTER2) -- node[midway,above]{$y_k$} (HD);
    \draw [draw,-latex] (ADC) --  node[right,text width=1.8em, align = right]{$Y$}(output);

    % Option (1): half-rectangle and channel on top on bottom line      
    \draw[dashed,thick,color=gray] ($(DAC.south)+(-0.85,1.5em)$) -- ($(DAC.south)+(-0.85,-2.7em)$);
    \draw[dashed,thick,color=gray] ($(ADC.south)+(0.75,1.5em)$) -- ($(ADC.south)+(0.75,-2.7em)$);
    \draw[dashed,thick,color=gray] ($(DAC.south)+(-0.85,-2.7em)$) -- ($(EO.south)+(-1.1em,-2.7em)$);
    \draw[dashed,thick,color=gray] ($(ADC.south)+(0.75,-2.7em)$) -- ($(EO.south)+(4.25,-2.7em)$);
    \node[right=0.8cm of DAC, yshift=-4.2em](){$Y=X+Z\sqrt{\sigma_{\thn}^2 + (X+\beta)^2\sigma_{\rin}^2}$};

    % Option (2): full-rectangle and channel below bottom line 
    % \draw[dashed,thick,color=gray] ($(DAC.south)+(-0.95,8em)$) -- ($(DAC.south)+(-0.95,-1.6em)$);
    % \draw[dashed,thick,color=gray] ($(DAC.south)+(-0.95,8em)$) -- ($(ADC.south)+(0.95,8em)$);
    % \draw[dashed,thick,color=gray] ($(ADC.south)+(0.95,8em)$) -- ($(ADC.south)+(0.95,-1.6em)$);
    % \draw[dashed,thick,color=gray] ($(DAC.south)+(-0.95,-1.6em)$) -- ($(ADC.south)+(0.95,-1.6em)$);
    % \node[right=1.3cm of DAC, yshift=-4.1em](){$Y=X+Z\sqrt{\sigma_{\mathrm{thn}}^2 + X^2\sigma_{\mathrm{rin}}^2}$};

    % Legend
    \node[above = 1.9cm of input, xshift=-1em] (elec_0){};
    \node[above = 1.9cm of input, xshift=1.5em] (elec_1){};
    \node[above = 1.55cm of input, xshift=-1em] (opt_0){};
    \node[above = 1.55cm of input, xshift=1.5em] (opt_1){};
    
    \draw[draw,-latex] (elec_0) -- node[right,text width=4.3em, align = right]{\footnotesize Electrical}(elec_1);
    \draw[draw,-latex,blue] (opt_0) -- node[right,text width=3.5em, align = right]{\footnotesize \textcolor{black}{Optical}}(opt_1);
 
\end{tikzpicture}
        \end{adjustbox}
    \end{subfigure}
    
    \caption{(Left) Received symbols distribution with signal-dependent RIN. (Right) IM-DD system under consideration.}
    \label{fig:IM-DD}
\end{figure}
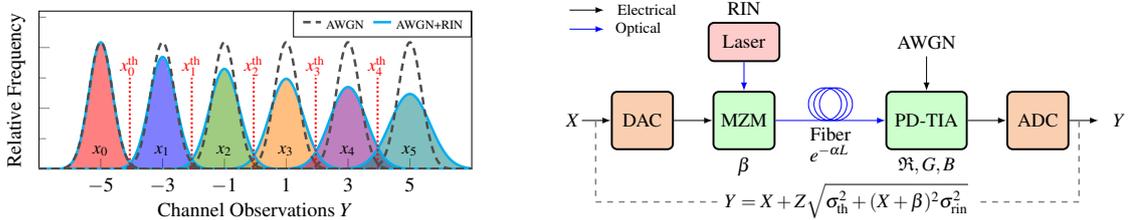

Since DCI target short distances, there is no optical amplification present in the link. Thus, when the transmitted optical power is relatively low, the dominant noise at the receiver is the thermal noise from the electronics. However, to achieve higher data rates, the optical power needs to be increased, which in turn results in the dominant noise to be the RIN from the transmitter laser \cite[Fig.~2]{szczerba20124}. RIN makes the variance of the laser intensity noise depend on the transmitted optical power level, leading to a different noise distribution for every PAM symbol. This effect is shown in Fig.~\ref{fig:IM-DD} (left). Compared to additive white Gaussian noise (AWGN), this signal-dependent noise complicates the digital signal processing in transceivers and degrades the performance. To compensate the effect of this signal-dependent noise, one can adjust the spacing between PAM symbols. This approach is referred to as geometric shaping (GS), and has been studied for PAM-4, e.g., in \cite{liang2023geometric}.

%Consequently, the main challenge for achieving these rates is to address the emerging signal-dependent noise caused by the RIN.

%Consequently, the main challenge for achieving these rates is to address the emerging signal-dependent noise caused by the increase in required optical power. In short-reach links without optical amplification, the dominant noise at the receiver is the thermal noise from the electronics, and at high optical powers, it is the RIN from the transmitter laser. This is illustrated in \cite[Fig.~2]{szczerba20124}.

In this paper, we study GS in a short-reach IM-DD optical link subject to RIN and thermal noise. Simulations are performed for different RIN parameters, where a PAM-6 constellation is optimized in order to minimize the symbol error rate (SER) in each case. To the best of our knowledge, we present, for the first time, the optimal geometrically-shaped PAM-6 constellations that lead to a SER improvement that relaxes the laser RIN parameter requirement by up to 3 dB.
%\vspace{-10pt}
% \cite[Fig.~4b]{che2023modulation}
\section{System Description and Equivalent Channel Model}

Following \cite[Sec. III]{che2021does}, we consider an IM-DD system without optical amplification, as shown in Fig.~\ref{fig:IM-DD} (right). The transmitter uses an O-band laser in continuous wave operation that is modulated by an external Mach-Zehnder modulator (MZM). We consider an MZM that is operated in push-pull configuration for chirp-free operation and is biased at its quadrature point for IM. The signal driving the modulator is the DAC output which contains the transmitted PAM-$M$ symbols $X$ taken from the vector $\mathbb{X}=(x_0, x_1,\dotsc, x_{M-1})=(\small{-(M\!-\!1),-(M\!-\!3),\dotsc, (M\!-\!1)})$. At the output of the MZM, the optical PAM levels will have an intensity noise variance proportional to $(X+\beta)^2$, where $\beta\geq |x_0|$ is a bias determined by the modulator dynamic extinction ratio (ER) \cite{che2023modulation}, and satisfies the constraint of non-negative IM of the system \cite{che2021does}. The signal then propagates through a standard single-mode fiber of length $L$ (km). We only consider the effect of signal attenuation, via the constant $\alpha$ (dB/km). Chromatic dispersion is ignored because of the operation in the zero dispersion regime, and short distances. To simulate variations in the received optical power, we vary the optical modulation amplitude (OMA) of the signal, which is defined as the swing between the maximum and minimum transmitted optical power.

At the receiver in Fig.~\ref{fig:IM-DD} (left), a single photodiode (PD) with responsivity $\mathfrak{R}$ is used to convert the incident optical power into an electrical current. During this conversion process, thermal noise modeled as AWGN is generated. The offset of the electrical current is removed, after which the resulting current is amplified and converted to a voltage by a transimpedance amplifier (TIA) with gain $G$. This gain is set to compensate for the electrical-to-optical conversion, fiber propagation, and optical-to-electrical conversion losses. Finally, the signal is sampled by the ADC to obtain the system output $Y$.

The entire IM-DD system can be modeled as a memoryless channel with additive noise:
\begin{equation} \label{eq:channel}
    Y = X + Z\sqrt{\sigma_\thn^2 + (X+\beta)^2\sigma_\rin^2}\;, \quad \textrm{with}\quad\! \sigma_\thn^2 = (G\cdot \mathrm{NEP})^2\cdot B, \quad \sigma_\rin^2 = 10^{\frac{\mathrm{RIN}}{10}}\cdot B,
\end{equation}
where $Z$ is a zero-mean and unit-variance Gaussian random variable. The variance contribution due to thermal noise $\sigma_\thn^2$, and the contribution due to the laser intensity noise $(X+\beta)^2\sigma_\rin^2$ are calculated from the PD-TIA noise equivalent power (NEP), and the laser RIN parameter, respectively \cite{szczerba20124}. We consider an effective system bandwidth of $B$, and thus, the variance is calculated from the noise power spectral density bandwidth limited to $B$. With this model, when a symbol $X=x_i$ is transmitted for $i = 0, 1,\dotsc, M\!-\!1$, the conditional probability density function of the channel $p_{Y|X}(y|x_i)$ has a Gaussian distribution with mean $x_i$ and variance $\sigma_i^2 = \sigma_\thn^2 + (x_i+\beta)^2\sigma_\rin^2$. The resulting received symbol distribution is shown in Fig.~\ref{fig:IM-DD} (left) where each variance $\sigma_i^2$ is signal-dependent. Note that when $\sigma_\rin^2=0$, an AWGN is obtained.
%This results in the received symbols distribution shown in Fig.~\ref{fig:IM-DD} (right) where each variance $\sigma_i^2$ is signal-dependent.

%Additionally, the symbol conditioned variance for this channel is given by .

\section{Geometric Shaping Optimization}

For PAM-$M$ signals and the channel in \eqref{eq:channel}, the average error probability is \cite{chagnon2014experimental}
\begin{equation} \label{eq:SER_analytical}
    \mathrm{SER}=\mathrm{Pr}\{\hat{X}\neq X\} = \frac{2}{M}\sum_{i=0}^{M-2}Q\left( \frac{x_{i+1}-x_i}{\sigma_{i+1}+\sigma_i} \right)\; , \quad \textrm{with}\;\ Q(x) = \frac{1}{\sqrt{2\pi}}\int_x^{\infty}e^{-\frac{u^2}{2}}\mathrm{d}u.
\end{equation}
This expression results from using the symbol hard-decision threshold $x_i^{\mathrm{th}} = (x_i\sigma_{i+1}+x_{i+1}\sigma_i)/(\sigma_i+\sigma_{i+1})$. 

For the IM-DD system under consideration, the swing of the MZM driving voltage is constrained in amplitude due to the modulator's finite driving range related with the ER. This constraint results in a limited peak-power launched into the fiber. This limitation is referred to as the peak-power constraint (PPC) in IM-DD systems with no optical amplification~\cite{che2023modulation}. Equivalently, due to the PPC, the two outer-most symbols of PAM-$M$ have a fixed magnitude. This leaves $M\!-\!2$ degrees of freedom for optimizing the positions of the inner symbols with GS. We now define the following constrained optimization problem 
\begin{equation} \label{eq:min_SER}
    \mathbb{X}^\star = \min_{x_1,\dotsc,x_{M-2}}\; \mathrm{SER}\; \quad \mathrm{s.t.}\;\ (x_0,x_{M-1}) =(-(M\!-\!1),(M-1)),\;\textrm{and }\ x_i<x_{i+1},\; \forall i. 
\end{equation}
The objective is to minimize the SER given by \eqref{eq:SER_analytical} over the constellation points $(x_1,...,x_{M-2})$, where the location of the remaining two constellation points, $x_0$ and $x_{M-1}$, are fixed by the PPC and ER. The constraints for $(x_0,x_{M-1})$ fix the minimum, and maximum power levels transmitted into the fiber, respectively. On the other hand, the inequality constraint in \eqref{eq:min_SER} defines the arrangement of the symbols within the PAM constellation. We study the SER since minimizing the SER maximizes the achievable information rate under symbol-wise hard-decision decoding~\cite[Eq. (7.15)]{handbookOptical}. We note that during simulations, we observed that maximizing the mutual information $I(X;Y)$ for this channel results in the same optimum constellations as minimizing the SER for large values of OMA. Thus, both optimization problems are equivalent for the range of OMAs that we are interested in for this application.

%To optimize \eqref{eq:min_SER} we employ \textcolor{red}{gradient-descent} to obtain the optimum constellation $\mathcal{X}^\star$ for different optical amplitude modulation (OMA) [CITA], and different RIN parameter values.

\section{Numerical Results}

We calculate the SER for GS PAM-$6$ via simulations. We choose PAM-6 as baseline since it has moderate bandwidth requirements to achieve $400$ Gbps. The parameters used during simulations are given in Table \ref{tab:sim_param}. The symbol rate is chosen such that a forward error correction (FEC) overhead of $12.9\%$ \cite{FEC200G} leads to a net bitrate close to $400$ Gbps. In the simulation setup, we utilize RIN values ranging from $-150$ to $-141$~dB/Hz. An additional case where $\mathrm{RIN}=-\infty$~dB/Hz is also included, which is equivalent to ignoring the laser intensity noise and only considering the thermal noise from the receiver, i.e., to the AWGN channel.

%\newpage
% \vspace{-8pt}
% \begin{figure}[H]
%     \centering

%     \begin{subfigure}[b]{0.48\textwidth}
%         \input{FigTikz/SER_PAM6_v2}
%         %\caption{}
%         \label{subfig:SER_RIN}
%     \end{subfigure}
%     ~\hspace{10pt}
%     \begin{subfigure}[b]{0.48\textwidth}
%         \input{FigTikz/C_PAM6_v2}
%         %\caption{}
%         \label{subfig:constellation}
%     \end{subfigure}
      
%     % \begin{subfigure}[b]{0.48\textwidth}
%     %     \input{FigTikz/GMI_PAM6_v2}
%     %     %\caption{}
%     %     \label{subfig:gmi}
%     % \end{subfigure}
    
%     \caption{(Left) Symbol error rate for PAM-$6$ with an equally-spaced constellation (dashed), geometrically-shaped constellation (solid). (Right) The optimized GS constellations.}
%     %\caption{The results from the simulations are shown in Fig.~2 (left) for four different RIN value scenarios, and OMA values.}
%     \label{fig:SER_PAM_large}
% \end{figure}

\vspace{-8pt}
\begin{figure}[H]
    \centering

    \begin{subfigure}[b]{0.48\textwidth}
        \definecolor{my_green}{rgb}{0.25, 0.6, 0.0}    %en la variables "my_green" guardo un color que he definido a partir de las componentes RGB. Es recomendable definir colores porque las librerías de tikZ no suelen tener mucha variedad de colores.
\definecolor{my_purple}{rgb}{0.5, 0, 0.5}

\newcommand{\lw}{0.6pt}
\newcommand{\lww}{0.8pt}
\newcommand{\mksc}{0.4}     %mark scale
\newcommand{\mkr}{*}     %mark scale

\newcommand{\graysolid}{\raisebox{2pt}{\tikz{\draw[color=black!50,solid,line width=\lw,mark=o,mark options={solid}](0,0) -- (2.5mm,0);}}}   %Símbolo de curva solida
\newcommand{\redsolid}{\raisebox{2pt}{\tikz{\draw[color=red,solid,line width=\lw,mark=o,mark options={solid}](0,0) -- (2.5mm,0);}}}   %Símbolo de curva solida
\newcommand{\bluesolid}{\raisebox{2pt}{\tikz{\draw[color=blue,solid,line width=\lw,mark=o,mark options={solid}](0,0) -- (2.5mm,0);}}}   %Símbolo de curva solida
\newcommand{\purplesolid}{\raisebox{2pt}{\tikz{\draw[color=my_purple,solid,line width=\lw,mark=o,mark options={solid}](0,0) -- (2.5mm,0);}}}   %Símbolo de curva 
\newcommand{\orangesolid}{\raisebox{2pt}{\tikz{\draw[color=orange,solid,line width=\lw,mark=o,mark options={solid}](0,0) -- (2.5mm,0);}}}   %Símbolo de curva solida
\newcommand{\greensolid}{\raisebox{2pt}{\tikz{\draw[color=my_green,solid,line width=\lw,mark=o,mark options={solid}](0,0) -- (2.5mm,0);}}}   %Símbolo de curva solida

\newcommand{\blacksdashed}{\raisebox{2pt}{\tikz{\draw[color=black,dashed,line width=\lw,mark=o,mark options={solid}](0,0) -- (2.5mm,0);}}}   %Símbolo de curva discontinua

\begin{tikzpicture}
    \begin{axis}[
    width=1.05\linewidth,  %ancho de la figura
    height=2.09in,
    ylabel={OMA [dBm]},
    xmin=-5.5, xmax=5.5,
    xtick={-5,-3,...,5},
    ymin=-6, ymax=6.2,
    ytick={-6,-4,...,6},
    xlabel={Optimized PAM Constellation $\mathbb{X}^\star$},
    ytick pos=left,
    xtick pos=bottom,
    y label style={yshift=-3pt},
    font=\footnotesize,
    grid=both,
    grid style = {densely dashed,lightgray!75},
    minor tick num=1,
    legend style = {legend pos=north east, font=\footnotesize, legend cell align=left, row sep=-0.5ex},
    %ytickten={-8,-6,...,0},
    %ytick={-8,-6,...,8},
    %xtick={2.7749, 4.7749, 6.7749, 8.7749},
    %xticklabels={$x_0$,$x_1$,$x_2$,$x_3$},
    ]

        %\pgfplotstableread[col sep=comma]{FigTikz/data_txt/Constellation_GS_PAM6_RIN-155to-140_NEP22_L2_ER5_Rb400_OH10.txt}\datatable
        \pgfplotstableread[col sep=comma]{FigTikz/data_txt/Constellation_GS_PAM6_RIN-150to-141_NEP22_L2_ER5_Rb400_OH12.9.txt}\datatable
        
        \node [font=\notsotiny,fill=white,anchor= north east] at (rel axis cs:0.96,0.94) {\shortstack[l]{
        RIN [\tiny{dB/Hz}]\\
        \hspace{0.015cm} \bluesolid \hspace{0.01cm} $-141$\\
        \hspace{0.015cm} \redsolid \hspace{0.01cm} $-144$\\
        %\purplesolid \hspace{0.01cm} $-150$\\
        \hspace{0.015cm} \greensolid \hspace{0.01cm} $-147$\\
        \hspace{0.015cm} \orangesolid \hspace{0.01cm} $-150$\\
        \hspace{0.015cm} \graysolid \hspace{0.01cm} $-\infty$
        }};
        
        % RIN -inf
        \foreach \col in {1,2,...,6} {
            % Add plots from file data, x-data is from column 1, y-data from column \col
            %\addplot[color=black!50, solid, line width=\lww, mark=\mkr, mark options={solid,scale=\mksc,fill=white}] table[x index=\col, y index=0] {\datatable};
            \addplot[color=black!50, solid, line width=\lww, mark=\mkr, mark options={solid,scale=\mksc,fill=white}] table[x index=\col, y index=0] {\datatable};
            %\addplot[color=black!50, solid, line width=\lww] table[x index=\col, y index=0] {\datatable};
        }
        % RIN -150
        \foreach \col in {7,8,...,12} {
            % Add plots from file data, x-data is from column 1, y-data from column \col
            \addplot[color=orange, solid, line width=\lww, mark=\mkr, mark options={solid,scale=\mksc,fill=white}] table[x index=\col, y index=0] {\datatable};
        }
        % RIN -147
        \foreach \col in {13,14,...,18} {
            % Add plots from file data, x-data is from column 1, y-data from column \col
            %\addplot[color=my_green, solid, line width=\lww, mark=\mkr, mark options={solid,scale=\mksc,fill=white}] table[x index=\col, y index=0] {\datatable};
            \addplot[color=my_green, solid, line width=\lww, mark=\mkr, mark options={solid,scale=\mksc,fill=white}] table[x index=\col, y index=0] {\datatable};
            %\addplot[color=my_green, solid, line width=\lww] table[x index=\col, y index=0] {\datatable};
        }
        % RIN -144
        \foreach \col in {19,20,...,24} {
            % Add plots from file data, x-data is from column 1, y-data from column \col
            %\addplot[color=red, solid, line width=\lww, mark=\mkr, mark options={solid,scale=\mksc,fill=white}] table[x index=\col, y index=0] {\datatable};
            \addplot[color=red, solid, line width=\lww, mark=\mkr, mark options={solid,scale=\mksc,fill=white}] table[x index=\col, y index=0] {\datatable};
            %\addplot[color=red, solid, line width=\lww] table[x index=\col, y index=0] {\datatable};
        }
        % RIN -141
        \foreach \col in {25,26,...,30} {
            % Add plots from file data, x-data is from column 1, y-data from column \col
            %\addplot[color=blue, solid, line width=\lww, mark=\mkr, mark options={solid,scale=\mksc,fill=white}] table[x index=\col, y index=0] {\datatable};
            \addplot[color=blue, solid, line width=\lww, mark=\mkr, mark options={solid,scale=\mksc,fill=white}] table[x index=\col, y index=0] {\datatable};
            %\addplot[color=blue, solid, line width=\lww] table[x index=\col, y index=0] {\datatable};
        }
        
        %\addlegendentry{$-140$};
        
        %\node [font=\footnotesize,anchor=north east] at (rel axis cs:0.98, 0.98) {\shortstack[l]{\textbf{(b)}}};

        %\draw[->] (axis cs:-7,0.00005) -- (axis cs:-0.5,0.00009);
        % \node [draw,font=\tiny,fill=white,anchor= south west] at (axis cs:-1.8,2e-3) {\shortstack[l]{
        % \bluesolid \hspace{0.01cm} $\mathrm{RIN}=-140$ dB/Hz\\
        % \redsolid \hspace{0.01cm} $\mathrm{RIN}=-145$ dB/Hz\\
        % \purplesolid \hspace{0.01cm} $\mathrm{RIN}=-150$ dB/Hz\\
        % \graysolid \hspace{0.01cm} $\mathrm{RIN}=-\infty$ dB/Hz
        % }};

        % Arrows and Deltas
        \node [font=\footnotesize] at (axis cs:-4,-5.2) {\shortstack[l]{
            $\Delta$}};
        \draw[line width=0.5pt, <->, >=latex] (axis cs:-5.0,-5.7) -- (-3.0,-5.7);
        
        \node [font=\footnotesize] at (axis cs:-2,-5.2) {\shortstack[l]{
            $\Delta$}};
        \draw[line width=0.5pt, <->, >=latex] (axis cs:-3.0,-5.7) -- (-1.0,-5.7);
        
        \node [font=\footnotesize] at (axis cs:0,-5.2) {\shortstack[l]{
            $\Delta$}};
        \draw[line width=0.5pt, <->, >=latex] (axis cs:-1.0,-5.7) -- (1.0,-5.7);

        \node [font=\footnotesize] at (axis cs:2,-5.2) {\shortstack[l]{
            $\Delta$}};
        \draw[line width=0.5pt, <->, >=latex] (axis cs:1.0,-5.7) -- (3.0,-5.7);
        
        \node [font=\footnotesize] at (axis cs:4,-5.2) {\shortstack[l]{
            $\Delta$}};
        \draw[line width=0.5pt, <->, >=latex] (axis cs:3.0,-5.7) -- (5.0,-5.7);

    \end{axis}

    \draw[line width=0.5pt, <->, >=latex] (0.3,3.72) -- (1.0,3.72);
    \node [font=\footnotesize] at (0.62,3.9) {\shortstack[l]{
            $0.6\Delta$}};
    
    \draw[line width=0.5pt, <->, >=latex] (1.05,3.72) -- (1.90,3.72);
    \node [font=\footnotesize] at (1.45,3.9) {\shortstack[l]{
            $0.75\Delta$}};
    
    \draw[line width=0.5pt, <->, >=latex] (1.95,3.72) -- (3.0,3.72);
    \node [font=\footnotesize] at (2.45,3.9) {\shortstack[l]{
            $0.95\Delta$}};
    
    \draw[line width=0.5pt, <->, >=latex] (3.05,3.72) -- (4.4,3.72);
    \node [font=\footnotesize] at (3.7,3.9) {\shortstack[l]{
            $1.2\Delta$}};
            
    \draw[line width=0.5pt, <->, >=latex] (4.45,3.72) -- (6.1,3.72);
    \node [font=\footnotesize] at (5.3,3.9) {\shortstack[l]{
            $1.5\Delta$}};

    % \newcommand{\blacksdotted}{\raisebox{2pt}{\tikz{\draw[color=black,dotted,line width=0.75pt,mark=o,mark options={solid}](0,0) -- (5mm,0);}}}   %Símbolo de curva 
    % \newcommand{\blackdiamond}{\raisebox{1pt}{\tikz{\node[draw,line width=1pt,scale=0.4,diamond,color=my_green,fill=white](){};}}}
    % \newcommand{\blackcircle}{\raisebox{1pt}{\tikz{\node[draw,line width=0.8pt,scale=0.5,circle,color=black,fill=white](){};}}}
    % \newcommand{\ksquare}{\raisebox{1pt}{\tikz{\node[draw,line width=1pt,scale=0.6,rectangle,color=red,fill=white](){};}}}
    
    % %\node [draw,font=\footnotesize,fill=white,anchor= south west] at (0.15,0.15) {\shortstack[l]{
    
    % \node[font=\footnotesize, color=red] at (6.5, 0.6) {\shortstack[l]{ PAM-4 }};
    % \node[font=\footnotesize, color=orange] at (6.5, 2.25) {\shortstack[l]{ PAM-6 }};
    % \node[font=\footnotesize, color=my_purple] at (6.5, 2.9) {\shortstack[l]{ PAM-8 }};
    %\node[] at (1.7,3.5) {\shortstack[l]{ $r_0^\star$ }};
    %\node[] at (3.3,3.5) {\shortstack[l]{ $r_1^\star$ }};
    %\node[] at (4.9,3.5) {\shortstack[l]{ $r_2^\star$ }};
\end{tikzpicture}
        %\caption{}
        \label{subfig:constellation}
    \end{subfigure}
    ~\hspace{-4pt}
    \begin{subfigure}[b]{0.48\textwidth}
        \definecolor{my_green}{rgb}{0.25, 0.6, 0.0}    %en la variables "my_green" guardo un color que he definido a partir de las componentes RGB. Es recomendable definir colores porque las librerías de tikZ no suelen tener mucha variedad de colores.
\definecolor{my_purple}{rgb}{0.5, 0, 0.5}

\definecolor{my_green2}{rgb}{0.0, 0.58, 0.0}

\definecolor{azure}{rgb}{0, 0.75, 0.90}
\definecolor{emerald}{rgb}{0.31, 0.78, 0.47}
\definecolor{lemon}{rgb}{1.0, 0.97, 0.0}

\newcommand{\lw}{0.8pt}
\newcommand{\lww}{0.8pt}
\newcommand{\mksc}{0.5}     %mark scale
\newcommand{\mkr}{*}     %mark 
\newcommand{\lses}{densely dotted}     % line style for Equally-Spaced results

\newcommand{\graysolid}{\raisebox{2pt}{\tikz{\draw[color=black!50,solid,line width=\lw,mark=o,mark options={solid}](0,0) -- (2.5mm,0);}}}   %Símbolo de curva solida
\newcommand{\redsolid}{\raisebox{2pt}{\tikz{\draw[color=red,solid,line width=\lw,mark=o,mark options={solid}](0,0) -- (2.5mm,0);}}}   %Símbolo de curva solida
\newcommand{\bluesolid}{\raisebox{2pt}{\tikz{\draw[color=blue,solid,line width=\lw,mark=o,mark options={solid}](0,0) -- (2.5mm,0);}}}   %Símbolo de curva solida
\newcommand{\purplesolid}{\raisebox{2pt}{\tikz{\draw[color=my_purple,solid,line width=\lw,mark=o,mark options={solid}](0,0) -- (2.5mm,0);}}}   %Símbolo de curva 
\newcommand{\orangesolid}{\raisebox{2pt}{\tikz{\draw[color=orange,solid,line width=\lw,mark=o,mark options={solid}](0,0) -- (2.5mm,0);}}}   %Símbolo de curva solida
\newcommand{\greensolid}{\raisebox{2pt}{\tikz{\draw[color=my_green,solid,line width=\lw,mark=o,mark options={solid}](0,0) -- (2.5mm,0);}}}   %Símbolo de curva solida

\newcommand{\blacksdashed}{\raisebox{2pt}{\tikz{\draw[color=black,\lses,line width=\lw,mark=o,mark options={solid}](0,0) -- (2.5mm,0);}}}   %Símbolo de curva 
\newcommand{\blacksolid}{\raisebox{2pt}{\tikz{\draw[color=black,solid,line width=\lw,mark=o,mark options={solid}](0,0) -- (2.5mm,0);}}}   %Símbolo de curva discontinua

\begin{tikzpicture}
    \begin{axis}[
    width=1.05\linewidth,  %ancho de la figura
    height=2.2in,
    ymode=log,
    xmin=-6, xmax=6,
    ymin=1e-12, ymax=1e-1,
    xlabel={OMA [dBm]},
    ylabel={Symbol Error Rate},
    ytick pos=left,
    xtick pos=bottom,
    font=\footnotesize,
    grid=both,
    grid style = {densely dashed,lightgray!75},
    minor tick num=1,
    yminorticks=true,
    %log basis y=10,
    legend style = {legend pos=north east, font=\footnotesize, legend cell align=left, row sep=-0.5ex},
    ytickten={-12,-10,...,0},
    xtick={-6,-4,...,6},
    yminorgrids,
    ]

        %\pgfplotstableread[col sep=comma]{FigTikz/data_txt/SER_PAM6_RIN-155to-140_NEP22_L2_ER5_Rb400_OH10.txt}\datatable
        \pgfplotstableread[col sep=comma]{FigTikz/data_txt/SER_PAM6_RIN-150to-141_NEP22_L2_ER5_Rb400_OH12.9.txt}\datatable

        \node [draw,font=\notsotiny,fill=white,anchor= north west] at (rel axis cs:0.015,0.65) {\shortstack[l]{
        RIN [\tiny{dB/Hz}]\\
        \hspace{0.015cm} \bluesolid \hspace{0.01cm} $-141$\\
        \hspace{0.015cm} \redsolid \hspace{0.01cm} $-144$\\
        %\purplesolid \hspace{0.01cm} $-150$\\
        \hspace{0.015cm} \greensolid \hspace{0.01cm} $-147$\\
        \hspace{0.015cm} \orangesolid \hspace{0.01cm} $-150$\\
        \hspace{0.015cm} \graysolid \hspace{0.01cm} $-\infty$
        }};

        % RIN -141
        \addplot[color=blue, \lses, line width=\lw, mark=\mkr, mark options={solid,scale=\mksc,fill=white}] table[x index=0,y index=9,col sep=comma] {\datatable};
        \addplot[color=blue, solid, line width=\lw, mark=\mkr, mark options={solid,scale=\mksc,fill=white}] table[x index=0,y index=10,col sep=comma] {\datatable};

        % RIN -144
        \addplot[color=red, \lses, line width=\lw, mark=\mkr, mark options={solid,scale=\mksc,fill=white}] table[x index=0,y index=7,col sep=comma] {\datatable};
        \addplot[color=red, solid, line width=\lw, mark=\mkr, mark options={solid,scale=\mksc,fill=white}] table[x index=0,y index=8,col sep=comma] {\datatable};

        % RIN -147
        \addplot[color=my_green, \lses, line width=\lw, mark=\mkr, mark options={solid,scale=\mksc,fill=white}] table[x index=0,y index=5,col sep=comma] {\datatable};
        \addplot[color=my_green, solid, line width=\lw, mark=\mkr, mark options={solid,scale=\mksc,fill=white}] table[x index=0,y index=6,col sep=comma] {\datatable};

        % RIN -150
        \addplot[color=orange, \lses, line width=\lw, mark=\mkr, mark options={solid,scale=\mksc,fill=white}] table[x index=0,y index=3,col sep=comma] {\datatable};
        \addplot[color=orange, solid, line width=\lw, mark=\mkr, mark options={solid,scale=\mksc,fill=white}] table[x index=0,y index=4,col sep=comma] {\datatable};

        % AWGN
        \addplot[color=black!50, solid, line width=\lw, mark=\mkr, mark options={solid,scale=\mksc,fill=white}] table[x={oma},y={awgn},col sep=comma] {\datatable};

        % \addplot[color=orange, solid, line width=\lw] table[x={oma},y={rin155}] {FigTikz/data_txt/SER_PAM6_RIN-155to-140_NEP22_L2_ER5_Rb400_OH10.txt};
        % \addplot[color=orange, dashed, line width=\lw] table[x={oma},y={rin155gs}] {FigTikz/data_txt/SER_PAM6_RIN-155to-140_NEP22_L2_ER5_Rb400_OH10.txt};

        %\addlegendentry{$-140$};
        
        %\node [font=\footnotesize,anchor=north west] at (rel axis cs:0.0,0.88) {\shortstack[l]{\textbf{(b)}}};

        %\draw[->] (axis cs:-7,0.00005) -- (axis cs:-0.5,0.00009);
        % \node [draw,font=\tiny,fill=white,anchor= south west] at (axis cs:-1.8,2e-3) {\shortstack[l]{
        % \bluesolid \hspace{0.01cm} $\mathrm{RIN}=-140$ dB/Hz\\
        % \redsolid \hspace{0.01cm} $\mathrm{RIN}=-145$ dB/Hz\\
        % \purplesolid \hspace{0.01cm} $\mathrm{RIN}=-150$ dB/Hz\\
        % \graysolid \hspace{0.01cm} $\mathrm{RIN}=-\infty$ dB/Hz
        % }};

        % Legend for Figure on Left
        % \node [draw,font=\notsotiny,fill=white,anchor= south west] at (rel axis cs:0.01,0.02) {\shortstack[l]{
        % \bluesolid \hspace{0.01cm} $\mathrm{RIN}=-141$ dB/Hz\\
        % \redsolid \hspace{0.01cm} $\mathrm{RIN}=-144$ dB/Hz\\
        % %\purplesolid \hspace{0.01cm} $\mathrm{RIN}=-150$ dB/Hz\\
        % \greensolid \hspace{0.01cm} $\mathrm{RIN}=-147$ dB/Hz\\
        % \orangesolid \hspace{0.01cm} $\mathrm{RIN}=-150$ dB/Hz\\
        % \graysolid \hspace{0.01cm} $\mathrm{RIN}=-\infty$  $\;\;\;\,$dB/Hz
        % }};

        % \node [draw,font=\notsotiny,fill=white,anchor= south west] at (rel axis cs:0.01,0.40) {\shortstack[l]{
        % \blacksdashed \hspace{0.01cm} Equally-Spaced\\
        % \blacksolid \hspace{0.01cm} Geometrically-Shaped
        % }};

        % Legend for Figure on Right

        \node [draw,font=\notsotiny,fill=white,anchor= south west] at (rel axis cs:0.015,0.02) {\shortstack[l]{
        \blacksdashed \hspace{0.01cm} Equally-Spaced\\
        \blacksolid \hspace{0.01cm} Geometrically-Shaped
        }};

        % Relevant Points (markers)
        \addplot[only marks, color=black!70, solid,line width=0.5pt, mark=diamond*, mark options={solid,scale=1.2,fill=lemon}] coordinates{(3, 1.5e-10)};
        \addplot[only marks, color=black!70, solid,line width=0.5pt, mark=diamond*, mark options={solid,scale=1.2,fill=lemon}] coordinates{(3.5, 3.5e-7)};
        %\addplot[only marks, color=black!70, solid,line width=0.5pt, mark=diamond*, mark options={solid,scale=1.2,fill=lemon}] coordinates{(6, 8e-5)};
        
        % Circles
        %\draw[] (axis cs:3.5, 5e-6) ellipse (3pt and 20pt);
        %\draw[] (axis cs:3, 1e-8) ellipse (3pt and 25pt);
        %\draw[->, >=latex, line width=0.8pt] (axis cs:3.5,1e-4) to[out=15, in=15] (3.5,5.5e-7);
        \draw[->, >=latex, line width=0.6pt, bend left=20] (axis cs:3.5,1e-4) to (3.5,6.2e-7); 
        \draw[->, >=latex, line width=0.6pt, bend left=20] (axis cs:3,7e-7) to (3,2.6e-10); 

        \node [font=\notsotiny,anchor= south east] at (rel axis cs:0.99,0.52) {\shortstack[c]{
            $3$ dB lower\\
            RIN
        }};
        
        \node [font=\notsotiny,anchor= south east] at (rel axis cs:0.95,0.28) {\shortstack[c]{
            $3$ dB lower\\
            RIN
        }};
        
    \end{axis}

    % \newcommand{\blacksdotted}{\raisebox{2pt}{\tikz{\draw[color=black,dotted,line width=0.75pt,mark=o,mark options={solid}](0,0) -- (5mm,0);}}}   %Símbolo de curva 
    % \newcommand{\blackdiamond}{\raisebox{1pt}{\tikz{\node[draw,line width=1pt,scale=0.4,diamond,color=my_green,fill=white](){};}}}
    % \newcommand{\blackcircle}{\raisebox{1pt}{\tikz{\node[draw,line width=0.8pt,scale=0.5,circle,color=black,fill=white](){};}}}
    % \newcommand{\ksquare}{\raisebox{1pt}{\tikz{\node[draw,line width=1pt,scale=0.6,rectangle,color=red,fill=white](){};}}}
    
    % %\node [draw,font=\footnotesize,fill=white,anchor= south west] at (0.15,0.15) {\shortstack[l]{
    
    % \node[font=\footnotesize, color=red] at (6.5, 0.6) {\shortstack[l]{ PAM-4 }};
    % \node[font=\footnotesize, color=orange] at (6.5, 2.25) {\shortstack[l]{ PAM-6 }};
    % \node[font=\footnotesize, color=my_purple] at (6.5, 2.9) {\shortstack[l]{ PAM-8 }};
    %\node[] at (1.7,3.5) {\shortstack[l]{ $r_0^\star$ }};
    %\node[] at (3.3,3.5) {\shortstack[l]{ $r_1^\star$ }};
    %\node[] at (4.9,3.5) {\shortstack[l]{ $r_2^\star$ }};
\end{tikzpicture}
        %\caption{}
        \label{subfig:SER_RIN}
    \end{subfigure}
        
    \caption{(Left) The optimized GS constellations $\mathbb{X}^\star$. (Right) Symbol error rate for PAM-$6$ with equally-spaced constellations (dotted), and geometrically-shaped constellations (solid). }
    %\caption{The results from the simulations are shown in Fig.~2 (left) for four different RIN value scenarios, and OMA values.}
    \label{fig:SER_PAM_large}
\end{figure}
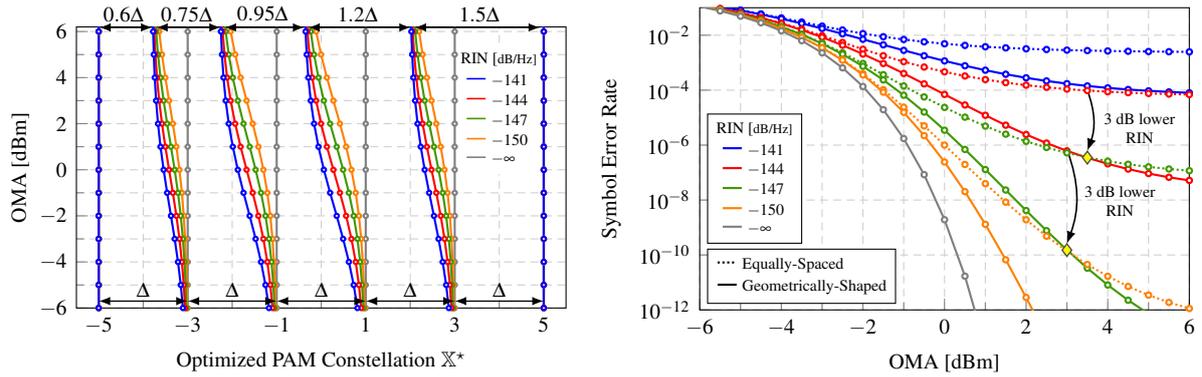

\begin{wraptable}[8]{r}{0.34\textwidth}
    \vspace{-16pt}
    \setlength{\columnsep}{5pt}
    \footnotesize
    \centering
    %\captionsetup{font=small}
    \caption{Simulation Parameters}
    \vspace{-3pt}
    \begin{tabular}{cc}
        \hline
        \textbf{System Parameter} & \textbf{Value}\\
        \hline
        Noise Bandwidth & $B = 105$ GHz\\
        Symbol Rate &   $R_s = 175$ GBd\\
        %Laser RIN   &    $\mathrm{RIN} \leq -141$ dB/Hz\\
        MZM ER  &   $\mathrm{ER}=5$ dB\\
        Fiber Attenuation   &   $\alpha=0.35$ dB/km\\
        Fiber Length    &   $L=2$ km\\
        PD Responsivity &   $\mathfrak{R}=0.5$ A/W\\
        PD-TIA NEP  &   $\mathrm{NEP}=22$ pA$/\sqrt{\textrm{Hz}}$\\
        \hline
    \end{tabular}
    \label{tab:sim_param}
\end{wraptable}

\vspace{-6pt}

The simulation results are shown in Fig.~\ref{fig:SER_PAM_large} for five different RIN values. The left figure shows the optimal constellation point locations for every RIN scenario as a function of OMA. Note that for the pure AWGN case, the optimal constellation corresponds to an equally-spaced (ES) constellation (shown in gray) due to the PPC. For the remaining scenarios, we observe that the optimal constellation is asymmetrical as the inner constellation points start to shift left of the ES constellation points as OMA increases. They then converge to the same spacing between symbols (where $\Delta$ is the spacing for the ES constellation) at asymptotically high OMA values. Intuitively, the optimal constellation points have lower optical power than that of the ES constellation, so that a smaller signal-dependent noise variance is experienced, see in Fig.~\ref{fig:IM-DD} (left).

Fig.~\ref{fig:SER_PAM_large} (right) shows that in this OMA range, by utilizing the optimized GS constellations, the SER improves by a few orders of magnitude with respect to the ES constellation in each RIN scenario. From these results, as OMA increases, the gap in SER between the GS and ES constellations also increases until the error floor is reached. With GS, the SER matches the error floor of a different system with a lower RIN parameter and ES constellation. This is evident for $\mathrm{OMA}\!\geq\! 3$ dBm, as denoted by the yellow markers in the figure. Here, the system with GS and RIN $-147$ dB/Hz matches the SER of the system with ES and RIN $-150$ dB/Hz. The same is true for the system with GS and RIN $-144$ dB/Hz, and ES with RIN $-147$ dB/Hz. This indicates a 3 dB relaxation on the RIN parameter of the laser. Thus, this suggest that by employing GS in the PAM-$6$ constellation, the laser design constraints can be relaxed depending on the target SER.

\section{Conclusions}

In this work, we explored the optimization of GS for an IM-DD link without optical amplification, under a PPC, and in the presence of both thermal noise and RIN. We showed via numerical simulations that an optimized geometrically-shaped PAM-6 constellation can significantly improve the SER of the system. This improvement allows for a laser design relaxation due to a higher RIN parameter tolerance of up to 3 dB, which is relevant for low-cost optical transmitters in DCI applications. 

% \section*{Acknowledgments}
% This research is part of the project Complexity-COnstrained LIght-coherent optical links (CoCoLi) funded by Holland High Tech $|$ TKI HSTM via the PPS allowance scheme for public-private partnerships.

\vspace{2mm}

\footnotesize
{
\noindent \textbf{Acknowledgements:}~This research is part of the project COmplexity-COnstrained LIght-coherent optical links (CoCoLi) funded by Holland High Tech $|$ TKI HSTM via the PPS allowance scheme for public-private partnerships.}

\bibliographystyle{IEEEtran}
\bibliography{IEEEabrv,references}

\begin{thebibliography}{1}
\providecommand{\url}[1]{#1}
\csname url@samestyle\endcsname
\providecommand{\newblock}{\relax}
\providecommand{\bibinfo}[2]{#2}
\providecommand{\BIBentrySTDinterwordspacing}{\spaceskip=0pt\relax}
\providecommand{\BIBentryALTinterwordstretchfactor}{4}
\providecommand{\BIBentryALTinterwordspacing}{\spaceskip=\fontdimen2\font plus
\BIBentryALTinterwordstretchfactor\fontdimen3\font minus \fontdimen4\font\relax}
\providecommand{\BIBforeignlanguage}[2]{{%
\expandafter\ifx\csname l@#1\endcsname\relax
\typeout{** WARNING: IEEEtran.bst: No hyphenation pattern has been}%
\typeout{** loaded for the language `#1'. Using the pattern for}%
\typeout{** the default language instead.}%
\else
\language=\csname l@#1\endcsname
\fi
#2}}
\providecommand{\BIBdecl}{\relax}
\BIBdecl

\bibitem{che2023modulation}
D.~Che and X.~Chen, ``Modulation format and digital signal processing for {IM-DD} optics at post-200{G} era,'' \emph{J. Lightw. Technol.}, vol.~42, no.~2, pp. 588--605, Jan. 2024.

\bibitem{hossain2021single}
M.~Sabbir-Bin~Hossain \emph{et~al.}, ``Single-{Lane} 402 {Gb/s} {PAM}-8 {IM/DD} transmission based on a single {DAC} and an {O-Band} commercial {EML},'' in \emph{Proc. Opto-Electron. and Commun. Conf. (OECC)}, Hong Kong, China, July 2021.

\bibitem{szczerba20124}
K.~Szczerba \emph{et~al.}, ``4-{PAM} for high-speed short-range optical communications,'' \emph{IEEE J. Opt. Commun. Netw.}, vol.~4, no.~11, pp. 885--894, Nov. 2012.

\bibitem{liang2023geometric}
E.~M. Liang and J.~M. Kahn, ``Geometric shaping for distortion-limited intensity modulation/direct detection data center links,'' \emph{IEEE Photon. J.}, vol.~15, no.~6, Dec. 2023.

\bibitem{che2021does}
D.~Che \emph{et~al.}, ``Does probabilistic constellation shaping benefit {IM-DD} systems without optical amplifiers?'' \emph{J. Lightw. Technol.}, vol.~39, no.~15, pp. 4997--5007, Aug. 2021.

\bibitem{chagnon2014experimental}
M.~Chagnon \emph{et~al.}, ``Experimental study of 112 {Gb/s} short reach transmission employing {PAM} formats and {SiP} intensity modulator at 1.3 $\mu$m,'' \emph{Opt. Express}, vol.~22, no.~17, pp. 21\,018--21\,036, Aug. 2014.

\bibitem{handbookOptical}
B.~Mukherjee \emph{et~al.}, \emph{Springer {H}andbook of {O}ptical {N}etworks}.\hskip 1em plus 0.5em minus 0.4em\relax Springer, 2020, ch.~7, p. 187.

\bibitem{FEC200G}
A.~Farhood \emph{et~al.}, ``{FEC Baseline Proposal for 200 Gbps per Lane IM-DD Optical PMDs},'' \emph{IEEE P802.3dj Task Force}, Mar. 2023.

\end{thebibliography}

% \begin{thebibliography}{99} %% use BibTeX or add references manually

% \bibitem{krishnan00} E. Krishnan, A. M. Shan, T. Rishi, L. A. Ajith, C. V.
% Radhakrishnan, \textit{On-line Tutorial on \LaTeX{}},
% ``Mathematics'' (Indian \TeX{} Users Group, 2000), \\
% \url{http://www.tug.org/tutorials/tugindia/chap11-scr.pdf}.

% \bibitem{vantrigt97} C. van Trigt, ``Visual system-response functions and estimating reflectance,''
% J. Opt. Soc. Am. A \textbf{14}, 741--755 (1997).

% \bibitem{masters93} T. Masters, \emph{Practical Neural Network Recipes in C++} (Academic, 1993).

% \bibitem{shoop97} B. L. Shoop, A. H. Sayles, and D. M. Litynski, ``New devices for optoelectronics: smart pixels,''
% in \emph{Handbook of Fiber Optic Data Communications},
% C. DeCusatis, D. Clement, E. Maass, and R. Lasky, eds. (Academic, 1997), pp. 705--758.

% \bibitem{kalman76} R. E. Kalman,``Algebraic aspects of the generalized inverse of a rectangular matrix,'' in
% \emph{Proceedings of Advanced Seminar on Generalized Inverse and Applications}, M. Z. Nashed, ed. (Academic, 1976), pp. 111--124.

% \bibitem{craig96} R. Craig and B. Gignac, ``High-power 980-nm pump lasers,''
% in \emph{Optical Fiber Communication Conference}, Vol. 2 of 1996 OSA Technical Digest Series (Optical Society of America, 1996), paper ThG1.

% \bibitem{steup96} D. Steup and J. Weinzierl, ``Resonant THz-meshes,''
% presented at the Fourth International Workshop on THz Electronics, Erlangen-Tennenlohe, Germany, 5--6 Sept. 1996.

% \end{thebibliography}

\end{document}